\documentclass[twocolumn]{autart}    
\usepackage[utf8]{inputenc}
\usepackage{comment}
\usepackage{algorithm}

\usepackage{algpseudocode}

\newcommand{\E}{\mathbb{E\,}}

\usepackage{graphicx}
\usepackage{amsmath}
\usepackage[version=4]{mhchem}
\usepackage{siunitx}
\usepackage{longtable,tabularx}
\newcommand{\R}{\mathbb{R}}
\usepackage{graphics} 
\usepackage{float}
\usepackage{epsfig} 
\usepackage{times} 
\usepackage{amssymb}  
\usepackage{mathtools}
\usepackage{color}
\usepackage[dvipsnames]{xcolor}
\usepackage{hyperref}

\newtheorem{lemma}{Lemma}
\newtheorem{theorem}{Theorem}
\newtheorem{corollary}{Corollary}

\usepackage{bm}
\usepackage{times}
\usepackage{soul}
\newcommand{\matlab}{\mbox{\textsc{Matlab}}}

\usepackage{graphicx}          

\begin{document}

\begin{frontmatter}
\title{Maximum Likelihood Recursive State Estimation using the Expectation Maximization Algorithm} 

\author[MR]{Mohammad S. Ramadan}\ead{msramada@eng.ucsd.edu},    
\author[RB]{Robert R. Bitmead}\ead{rbitmead@eng.ucsd.edu} 

\address[MR]{Department of Mechanical \&\ Aerospace Engineering, University of California, San Diego, La Jolla CA 92093-0411, USA, and Electrical \&\ Computer Engineering Department, San Diego State University, San Diego CA 92182, USA}  
\address[RB]{Department of Mechanical \&\ Aerospace Engineering, University of California, San Diego, La Jolla CA 92093-0411, USA.}             


\begin{abstract}
A Maximum Likelihood recursive state estimator is derived for non-linear and non-Gaussian state-space models. The estimator combines a particle filter to generate the conditional density and the Expectation Maximization algorithm to compute the maximum likelihood state estimate iteratively. Algorithms for maximum likelihood state filtering, prediction and smoothing are presented. The convergence properties of these algorithms, which are inherited from the Expectation Maximization algorithm, are proven and examined in two examples. It is shown that, with randomized reinitialization, which is feasible because of the algorithm simplicity, these methods are able to converge to the Maximum Likelihood Estimate (MLE) of multimodal, truncated and skewed densities, as well as those of disjoint support.
\end{abstract}

\end{frontmatter}

\section{Introduction}
We derive a maximum likelihood recursive state estimator for nonlinear systems based on: knowledge of the system equations and process noise and measurement noise densities; the operation of a particle filter to propagate conditional densities; and the Expectation Maximization (EM) algorithm \cite{dempster1977maximum,moon1996expectation}. If the state at time $k$ is denoted $\zeta_k$, and given the sequence of outputs up to time $k$, namely $\mathcal Y_{0:k}=\{y_0,y_1,\dots,y_k\}$, the algorithm provides an estimate of the maximizer of the posterior state density $p(\zeta_k\vert \mathcal Y_{0:k})$. This contrasts with Least-Squares estimators, which produce the conditional mean of this density. The contribution of the paper is to provide an easily implementable algorithm which is recursive and which yields this Maximum Likelihood Estimate (MLE). We propagate the conditional densities, both prior and posterior, using a particle filter, and the EM algorithm is used to compute the MLE of the posterior density from the particles of the prior density, the current measurement and the known noise densities. The observation is that knowledge of the noise densities permits computation of the gradient of successive functions whose maximizers are non-decreasing in their likelihood with respect to the posterior density, which follows from the desirable convergence properties of the EM algorithm. If the system under consideration has Gaussian additive noises and linear measurement equation, these successive functions are quadratic and the algorithm simplifies into a fixed-point iteration. The paper clarifies these details in relation to the particle filter, the EM algorithm and MLE.

 In many scenarios, the conditional state density can: be multi-modal as when tracking a group of objects \cite{mihaylova2014overview,septier2009tracking}; have skewness, limited support, one-sidedness as in some Stochastic Volatility Models \cite{vankov2019filtering,li2020leverage}; or be governed by inequality state constraints \cite{li2016auxiliary}, leading to truncated or disjoint support such as might arise in stochastic model Predictive Control. The justification for the choice of the conditional mean as the point estimate of a filter may be weakened or  render an estimate which is infeasible under the density. In such scenarios, the MLE may have more reasonable grounds for use. Since finding the MLE requires optimization over the posterior density, the recursive procedure involves two stages: the propagation of the posterior density, for which we rely on a particle filter; and, an iterative optimization to find the maximizer. However, the density provided by the particle filter is a collection of point masses, usually with the same singleton mass, and so is not suited to Newton-like optimization. A novelty of this paper and its algorithms is that the previous time's density plus the current measurement are combined with the known continuous densities of the process and measurement noises to yield a gradient based maximizer. Effectively, this procedure yields a smooth interpolation of the particle filter's posterior density and of its gradient.
 
In \cite{malik2011particle}, a piecewise continuous cumulative distribution function is constructed from the weighted particles approximating the filtered state density $p(\zeta_k|\mathcal{Y}_{0:k})$, limited to the case when the state-space is one-dimensional. In \cite{dejong2013efficient}, this solution is extended to a multi-dimensional state-space. However, it has a computational complexity of $\mathcal{O}(N^2)$, where $N$ is the number of particles, and a nonstandard particle filtering scheme, as reported in \cite{kantas2015particle}. Earlier, in the construction of the auxiliary particle filter  \cite{pitt1999filtering}, it was shown that it is possible to construct a smooth approximation of $p(\zeta_k|\mathcal{Y}_{0:k})$ under the name of ``the empirical filtered density'' using the weighted particles approximating $p(\zeta_{k-1}|\mathcal{Y}_{0:k-1})$ and the current measurement.We extend this construction with the computation of an approximate gradient, as shall be explained shortly. This constructed smooth density could be directly maximized using various search methods and the new calculation permits gradient search for the MLE. We introduce the EM algorithm as an iterative procedure to achieve this, perhaps with multiple initial starting values in the multimodal case. This yields a sequence of estimates with non-decreasing likelihood which converges to a stationary point \cite{wu1983convergence,dempster1977maximum}. This will be supported by computational examples in Sections \ref{LGSSM} and \ref{section2E}. In effect, the EM algorithm smooths and recenters the empirical filtered density at the E-step before maximizing at the M-step. The iterations refine and localize the smoothing to approach the MLE. Ultimately, this is limited by the quality of the empirical density generated by the particle filter. Without such smoothing, it is not possible to find the maximizer of the particle filter's point-mass density.

The EM algorithm \cite{dempster1977maximum} is  versatile with many variations \cite{meng1993maximum,wei1990monte,lange1995gradient} and applications primarily in parameter estimation. It consists of two major steps: the expectation (E) step, in which an approximant to the log-likelihood function is constructed, followed by the maximization (M) step, which seeks to find the maximizer of this function. Hence, EM iterations deliver a sequence of functions for which their maximizers are non-decreasing with respect to the likelihood function under consideration \cite{wu1983convergence}. Equipped with Monte-Carlo particle methods, the EM algorithm is widely used in parameter estimation for nonlinear non-Gaussian systems \cite{schon2011system,poyiadjis2011particle,lindholm2018learning,le2013convergence}. We examine the use of EM in the context of state estimation; to our knowledge this is a novel application and different in principle, since, unlike parameters, the components of the state are dynamically changing according to the tailored system model.

The particle filter \cite{doucet2000sequential}, itself, is a common feature of nonlinear estimation and operates by propagating the conditional densities as characterized by a collection of samples, called particles, distributed with this density. For a basic particle filter, the time update pass, i.e. posterior-to-prior densities, uses the state equation and samples from the process noise density. This is followed by the measurement update, prior-to-posterior densities, relying on weighting and possibly resampling these particles based on the output equation and the output noise density.

In this paper, we use the weighted particles' approximation of $p(\zeta_{k-1}|\mathcal{Y}_{0:k-1})$ and the current observation $y_k$ with the EM algorithm to construct a sequence of functions of $\zeta_k$, whose maximizers are non-decreasing in their likelihood with respect to $p(\zeta_k|\mathcal{Y}_{0:k})$. In Section~\ref{section2}, the underlying assumptions on the state-space model under consideration are elucidated. In Section~\ref{section3}, the EM state filtering (EMSF) algorithm is introduced, and implemented in two examples \ref{LGSSM} and \ref{section2E}. First, it is applied to state filtering of a linear Gaussian state-space model, and since the conditional mean is equivalent to the conditional mode of such systems \cite{rauch1965maximum}, it is compared to the Kalman filter solution, and shown to be convergent to the conditional mode (the global maximizer) from a random initial guess. This establishes the efficacy of the algorithm and permits statements concerning its complexity. In the second example, it is used to track the MLE of a multi-modal state distribution of a nonlinear state-space model. It is also shown, in Section~\ref{section2F} that this new EM state filter (EMSF) algorithm reduces to a convergent fixed-point iteration method under mild assumption. The extensions of this algorithm to the contexts of state prediction and smoothing are provided in Sections~\ref{section4} and \ref{section5}.

\section{Problem Statement} \label{section2}
 Consider the following nonlinear dynamical system affine in the noises\footnote{The notation $\sim$ here means ``is distributed as the density on the right''.}:
 \begin{subequations}
    \begin{align}
    \zeta_{k+1}&=f(\zeta_k)+w_k, \, \zeta_k \in \R^{r_\zeta},\label{NonSysa}\\
    y_k&=g(\zeta_k)+v_k, \, y_k \in \R^{r_y}, \label{NonSysb}\\
    w_k &\sim W_k(.),\,
    v_k \sim V_k(.),\nonumber\\
    \zeta_0 &\sim p_0(.) \nonumber
    \end{align}\label{NonSys}
    \end{subequations}
where,
\begin{itemize}
    \item the stochastic processes $\{w_k\}$ and $\{v_k\}$ are white and independent, and their probability density functions, $W_k(\cdot)$ and $V_k(\cdot)$ are known and continuously differentiable for all $k$,
    \item $\zeta_k$ is the state vector, with initial value, $\zeta_0$, having a known initial distribution which is independent of $\{w_k\}$ and $\{v_k\}$,
    \item $y_k$ is the measurement vector,
    \item the functions $f$ and $g$ are known continuously differentiable functions.
\end{itemize}
Given the observation sequence, $\mathcal{Y}_{0:k}=\{y_0,y_1,\dots,y_{k}\}$, the problem is to find the MLE of the state-vector $\zeta_k$ using the log-likelihood function $\log p(\zeta_k | \mathcal{Y}_{0:m})$, for $m=k$ (filtering), $m>k$ (smoothing), and $m<k$ (prediction).

\section{ML state filtering by the expectation maximization algorithm}\label{section3}
Given, at time $k$:
\begin{itemize}
    \item \vskip -3mm the problem statement and assumptions as in Section~\ref{section2},
    \item the measurement $y_k$,
    \item the filtered density at time $k-1$, i.e. $p(\zeta_{k-1}|\mathcal{Y}_{0:k-1})$,
    \item initial time-$k$ state estimate $\zeta^0_k$,
\end{itemize}
\vskip -3mm we set iteration counter $i=0$ and seek  to find $\zeta_k^{i+1}$ such that $\log p(\zeta_k^{i+1} | \mathcal{Y}_{0:k}) \geq\log p(\zeta_k^i | \mathcal{Y}_{0:k})$ \cite{wu1983convergence}. This is achieved by two major steps: the expectation step (E-step), followed by the maximization step (M-step). We continue with these iterations in $i$ until we satisfy a convergence criterion, which is guaranteed by the non-decreasing property. At this stage, we either restart with a different initial condition, $\zeta_k^0$, or move to time $k+1$.
\subsection{E-step}
The E-step is performed by evaluating the following expectation,
\begin{align}
     Q_\zeta(\zeta_k,\zeta_k^i)&=\E_{\zeta_{k-1}|\zeta_k^{i},\mathcal{Y}_{0:k}} \{ \log p(\zeta_k|\zeta_{k-1},\mathcal{Y}_{0:k})|\zeta_k^{i},\mathcal{Y}_{0:k}\}\nonumber\\
     &\hskip -15mm =\int \log p(\zeta_k|\zeta_{k-1},\mathcal{Y}_{0:k})p(\zeta_{k-1}|\zeta_k^{i},\mathcal{Y}_{0:k})\,d\zeta_{k-1} \label{Q_int}
\end{align}
whose calculation in terms of available densities will be presented shortly in the following theorem. 

Before presenting the theorem, we first derive a lemma detailing a conditional independence result, which will be used repeatedly in the proofs and derivations.
\vskip 3mm
\begin{lemma} \label{lemma1}
For system \eqref{NonSys} at time $n\geq k$, $y_n$ conditioned on $\zeta_k$ is conditionally independent from both $\zeta_{0:k-1}=\{\zeta_0,\zeta_1,\hdots,\zeta_{k-1}\}$ and $\mathcal{Y}_{0:k-1}$, that is
\begin{align*}
    p(y_n|\zeta_k,\zeta_{0:k-1},\mathcal{Y}_{0:k-1})=p(y_n|\zeta_k)
\end{align*}
\end{lemma}
\textit{Proof.}
From \eqref{NonSys} and for $n\geq k$, $y_n$ is a function of $\zeta_k$, $ w_{k:n-1}$ and $v_n$. Since $w_{k:n-1}$ and $v_n$ are independent from $\zeta_{0:k-1}$ and $\mathcal Y_{0:k-1}$, we have the result. \hfill $\square$
\vskip 3mm
\begin{theorem}\label{theorem1}
Given the dynamical system (\ref{NonSys}), the filtered density $p(\zeta_{k-1}|\mathcal{Y}_{0:k-1})$, the current measurement $y_k$, and $\zeta_k^i$, the integral in (\ref{Q_int}) is, apart from a positive multiplicative constant and an additive constant both independent from $\zeta_k$, equal to
\begin{align}
    &Q_\zeta(\zeta_k,\zeta_k^i)=\nonumber \\
    &\int \Big [\log V_k(y_k-g(\zeta_k))+ \log W_{k-1}(\zeta_k-f(\zeta_{k-1})) \Big ]\times \nonumber \\&W_{k-1}(\zeta_k^i-f(\zeta_{k-1}))p(\zeta_{k-1}|\mathcal{Y}_{0:k-1})\,d\zeta_{k-1},
\end{align}
and its gradient is given by
\begin{align}
     \nabla_{\zeta_k} Q_\zeta(\zeta_k,\zeta_k^i)
     &=\int \Big [\nabla_{\zeta_k}\log V_k(y_k-g(\zeta_k))+ \nonumber\\&\hskip -10mm \nabla_{\zeta_k}\log W_{k-1}(\zeta_k-f(\zeta_{k-1})) \Big ]\times \nonumber \\&\hskip -10mm W_{k-1}(\zeta_k^i-f(\zeta_{k-1}))p(\zeta_{k-1}|\mathcal{Y}_{0:k-1}) d\zeta_{k-1}. \label{Q_int_grad}
\end{align}
\end{theorem}
\textit{Proof.}~
We use Bayes rule and Lemma~\ref{lemma1}, to yield
\begin{align*}
   p(\zeta_k|\zeta_{k-1},\mathcal{Y}_{0:k})&= \frac{p(y_k|\zeta_k,\zeta_{k-1},\mathcal{Y}_{0:k-1})p(\zeta_k|\zeta_{k-1},\mathcal{Y}_{0:k-1})}{p(y_k|\zeta_{k-1},\mathcal{Y}_{0:k-1})} \nonumber\\
   &= \frac{p(y_k|\zeta_k)p(\zeta_k|\zeta_{k-1})}{p(y_k|\zeta_{k-1})}.
\end{align*}
Thus,
\begin{align}
  \log  p(\zeta_k|\zeta_{k-1},\mathcal{Y}_{0:k})&= \log p(y_k|\zeta_k)+\log p(\zeta_k|\zeta_{k-1})\nonumber \\& \hskip -10mm -\log p(y_k|\zeta_{k-1}) \nonumber\\
  &=\log V_k(y_k-g(\zeta_k)) \nonumber \\&\hskip -10mm +\log W_{k-1}(\zeta_k-f(\zeta_{k-1}))-\tilde C \label{term1}
\end{align}
where $\tilde C=\log p(y_k|\zeta_{k-1})$ is independent from $\zeta_k$ and so integrates to an additive constant in the expectation (\ref{Q_int}).

Similarly, by Bayes rule and Lemma~\ref{lemma1},
\begin{align*}
 p(\zeta_{k-1}|\zeta_k^{i},\mathcal{Y}_{0:k})&=p(\zeta_{k-1}|\zeta_k^{i},\mathcal{Y}_{0:k-1},y_k),\\
 &=\frac{p(y_k|\zeta_k^i,\zeta_{k-1},\mathcal{Y}_{0:k-1})p(\zeta_{k-1}|\zeta_k^{i},\mathcal{Y}_{0:k-1})}{p(y_k|\zeta_k^i,\mathcal{Y}_{0:k-1})},\\
 &=\frac{p(y_k|\zeta_k^i)p(\zeta_{k-1}|\zeta_k^{i},\mathcal{Y}_{0:k-1})}{p(y_k|\zeta_k^i)},\\
 &=p(\zeta_{k-1}|\zeta_k^{i},\mathcal{Y}_{0:k-1}),
\end{align*}
and, applying Bayes rule again,
\begin{align*}
p(\zeta_{k-1}|\zeta_k^{i},\mathcal{Y}_{0:k})       &=\frac{p(\zeta_k^i|\zeta_{k-1})p(\zeta_{k-1}|\mathcal{Y}_{0:k-1})}{p(\zeta_{k}^i|\mathcal{Y}_{0:k-1})}. 
\end{align*}
Hence,
\begin{align}
    p(\zeta_{k-1}|\zeta_k^{i},\mathcal{Y}_{0:k})&= \nonumber \\& \hskip -5mm\frac{W_{k-1}(\zeta_k^i-f(\zeta_{k-1}))p(\zeta_{k-1}|\mathcal{Y}_{0:k-1})}{\hat C} \label{term2}
\end{align}
where $\hat C=p(\zeta_{k}^i|\mathcal{Y}_{0:k-1})$ is independent from $\zeta_{k-1}$ and leads to a positive multiplicative constant in the expectation (\ref{Q_int}).

Substituting (\ref{term1}) and (\ref{term2}) into (\ref{Q_int}) and neglecting the additive and positive multiplicative constants above yields
\begin{align*}
    Q_\zeta(\zeta_k,\zeta_k^i)&=\int \Big\{\big [\log p(y_k|\zeta_k)+\log p(\zeta_k|\zeta_{k-1}) \big ]\times\\&\hskip 15mm p(\zeta_k^i|\zeta_{k-1})p(\zeta_{k-1}|\mathcal{Y}_{0:k-1})\Big\}\, d\zeta_{k-1} \\
    &\hskip -15mm =\int \Big\{\big [\log V_k(y_k-g(\zeta_k))+ \log W_{k-1}(\zeta_k-f(\zeta_{k-1})) \big ]\times  \\&W_{k-1}(\zeta_k^i-f(\zeta_{k-1}))p(\zeta_{k-1}|\mathcal{Y}_{0:k-1}) \Big\}\,d\zeta_{k-1}.
\end{align*}
The smoothness assumptions in Section~\ref{section2} permit exchanging differentiation and integration to arrive at the gradient formula.
\begin{align*}
     \nabla_{\zeta_k} Q_\zeta(\zeta_k,\zeta_k^i)
     &=\nabla_{\zeta_k}\int \Big\{\big [\log V_k(y_k-g(\zeta_k))+\\&\hskip -10mm\log W_{k-1}(\zeta_k-f(\zeta_{k-1})) \big ]\times \nonumber \\&\hskip -10mmW_{k-1}(\zeta_k^i-f(\zeta_{k-1}))p(\zeta_{k-1}|\mathcal{Y}_{0:k-1})\Big\}\, d\zeta_{k-1}\\
     &=\int \Big\{\big [\nabla_{\zeta_k}\log V_k(y_k-g(\zeta_k)) \nonumber\\&\hskip -10mm+\nabla_{\zeta_k}\log W_{k-1}(\zeta_k-f(\zeta_{k-1})) \big ]\times \nonumber \\&\hskip -10mmW_{k-1}(\zeta_k^i-f(\zeta_{k-1}))p(\zeta_{k-1}|\mathcal{Y}_{0:k-1})\Big\}\, d\zeta_{k-1}.
\end{align*}
\hfill $\square$

The gradient $\nabla_{\zeta_k} Q_\zeta(\zeta_k,\zeta_k^i)$ can be used within a gradient search algorithm to evaluate the M-step, which follows immediately.

\subsection{M-step}
The M-step maximizes the expectation in the E-step w.r.t $\zeta_k$, choosing $\zeta_k^{i+1}$ to be the maximizer,
\begin{align}
    \zeta_k^{i+1}=\text{arg}\max_{\zeta_k \in \R^{r_\zeta}} Q(\zeta_k,\zeta_k^i) \label{Mstep.}
\end{align}
This choice of $\zeta_k^{i+1}$ will be shown in Theorem~\ref{Theorem2} to be non-decreasing in its likelihood compared to $\zeta_k^i$. A proof of the general statement of EM algorithm can be found in \cite{wu1983convergence}. We present the proof in the context of state filtering.

\vskip 3mm
\begin{theorem}
The sequence $\{\zeta_k^i\}_{i\geq0}$, commencing at $\zeta_k^0 \in \R^{r_\zeta}$ and satisfying the recursion (\ref{Mstep.}), has the property that $\{ \log p(\zeta_k^i | \mathcal{Y}_{0:k})\}_{i \geq 0}$ is monotonically non-decreasing. \label{Theorem2}
\end{theorem}
\textit{Proof.}~
The log-likelihood function $\log p(\zeta_k | \mathcal{Y}_{0:k})$ can be expressed as
\begin{align*}
     \log p(\zeta_k|\mathcal{Y}_{0:k})&=\log p(\zeta_k,\zeta_{k-1}|\mathcal{Y}_{0:k})-\log p(\zeta_{k-1}|\mathcal{Y}_{0:k},\zeta_{k})\nonumber\\
     &=\log p(\zeta_k|\zeta_{k-1},\mathcal{Y}_{0:k})+ \log p(\zeta_{k-1}|\mathcal{Y}_{0:k})\nonumber\\&\hskip 5mm -\log p(\zeta_{k-1}|\mathcal{Y}_{0:k},\zeta_{k}).
\end{align*}
Taking the expectation with $p(\zeta_{k-1}\vert\zeta_k^i,\mathcal{Y}_{0:k})$ as in \eqref{Q_int},
\begin{align*}
     \log p(\zeta_k|\mathcal{Y}_{0:k})
     &=Q_\zeta(\zeta_k,\zeta_k^i)+ \mathbb{E}_{\zeta_{k-1}|\zeta_k^i,\mathcal{Y}_{0:k}}\{\log p(\zeta_{k-1}|\mathcal{Y}_{0:k})\}\nonumber\\&\hskip 5mm -\mathbb{E}_{\zeta_{k-1}|\zeta_k^i,\mathcal{Y}_{0:k}}\{\log p(\zeta_{k-1}|\mathcal{Y}_{0:k},\zeta_{k})\}.
\end{align*}
Hence, the log-likelihood of $\zeta_k^i$,
\begin{align*}
     \log p(\zeta_k^i|\mathcal{Y}_{0:k})
     &=Q_\zeta(\zeta_k^i,\zeta_k^i)+ \mathbb{E}_{\zeta_{k-1}|\zeta_k^i,\mathcal{Y}_{0:k}}\{\log p(\zeta_{k-1}|\mathcal{Y}_{0:k})\}\nonumber\\&
     \hskip 5mm -\mathbb{E}_{\zeta_{k-1}|\zeta_k^i,\mathcal{Y}_{0:k}}\{\log p(\zeta_{k-1}|\mathcal{Y}_{0:k},\zeta_{k}^i)\}.
\end{align*}
Subtracting these two equations,
\begin{align*}
     &\log p(\zeta_k|\mathcal{Y}_{0:k}) - \log p(\zeta_k^i|\mathcal{Y}_{0:k})=Q_\zeta(\zeta_k,\zeta_k^i)-Q_\zeta(\zeta_k^i,\zeta_k^i)\nonumber\\
     &+\mathbb{E}_{\zeta_{k-1}|\zeta_k^i,\mathcal{Y}_{0:k}}\{\log p(\zeta_{k-1}|\mathcal{Y}_{0:k},\zeta_{k}^i)-\log p(\zeta_{k-1}|\mathcal{Y}_{0:k},\zeta_{k})\}
\end{align*}
where the term,
\begin{align*}
    &\mathbb{E}_{\zeta_{k-1}|\zeta_k^i,\mathcal{Y}_{0:k}}\{\log p(\zeta_{k-1}|\mathcal{Y}_{0:k},\zeta_{k}^i)-\log p(\zeta_{k-1}|\mathcal{Y}_{0:k},\zeta_{k})\}\nonumber\\
    &=\int _{\R^{r_\zeta}} \frac{\log p(\zeta_{k-1}|\mathcal{Y}_{0:k},\zeta_{k}^i)}{\log p(\zeta_{k-1}|\mathcal{Y}_{0:k},\zeta_{k})}p(\zeta_{k-1}|\zeta_k^i,\mathcal{Y}_{0:k})d\zeta_{k-1} \nonumber\\
    &\geq 0.
\end{align*}
Since the right-hand side is the Kullback-Leibler divergence metric between these two densities, this quantity is non-negative. (See Example 7.7 in \cite{lehmann2006theory}.) Thus,
\begin{align*}
    \log p(\zeta_k|\mathcal{Y}_{0:k}) - \log p(\zeta_k^i|\mathcal{Y}_{0:k}) \geq Q_\zeta(\zeta_k,\zeta_k^i)-Q_\zeta(\zeta_k^i,\zeta_k^i).
\end{align*}
It follows from the definition of $\zeta_k^{i+1}$ as the maximizer of $Q_\zeta(\zeta_k,\zeta_k^i)$ that
\begin{align*}
    Q_\zeta(\zeta_k^{i+1},\zeta_k^i)-Q_\zeta(\zeta_k^i,\zeta_k^i) \geq 0.
\end{align*}
Therefore,
\begin{align*}
    \log p(\zeta_k^{i+1}|\mathcal{Y}_{0:k}) - \log p(\zeta_k^i|\mathcal{Y}_{0:k}) \geq 0,
\end{align*}
which completes the proof.
\hfill $\square$
\begin{corollary} If the log-likelihood function $\log p(\zeta_k|\mathcal{Y}_{0:k})$ is unimodal with the MLE being the only stationary point, then $\{\zeta_k^i\}_{i\geq0}$ converges to this MLE. \label{corollary1}
\end{corollary}
This parallels Corollary 1 in \cite{wu1983convergence}.

The EM algorithm for ML state filtering at time-step $k$ follows.
\begin{algorithm}[H]
\caption{ }\label{algorithm1}
\begin{algorithmic}[1]
        \State Given the current measurement $y_k$, the prior filtered density $p(\zeta_{k-1}|\mathcal{Y}_{0:k-1})$ and prior MLE $\zeta_{k-1}^*$, set $i \gets 0$ and let $\zeta_k^0=\zeta_k^i=f(\zeta_{k-1}^*) + \mathbb{E}w_k$;
        \State Evaluate the integral:
        \begin{align}
             Q_\zeta(\zeta_k,\zeta_k^i)&=\int\Big\{ \Big [\log V_k(y_k-g(\zeta_k))+\nonumber \\& \hskip 15mm\log W_{k-1}(\zeta_k-f(\zeta_{k-1})) \Big ]\times \nonumber \\& \hskip -10mm W_{k-1}(\zeta_k^i-f(\zeta_{k-1}))p(\zeta_{k-1}|\mathcal{Y}_{0:k-1})\Big\}\, d\zeta_{k-1} \label{Q_int_Alg}
        \end{align}
        \State Evaluate $\zeta_k^{i+1}$:
        \begin{equation}
            \zeta_k^{i+1} =\text{arg} \max_{\zeta_k \in \R^{r_\zeta}} Q_\zeta(\zeta_k,\zeta_k^{i})
        \end{equation}
        \State If $\{\zeta_k^i\}_{i\geq0}$ satisfies a convergence criterion: set $\zeta_k^*=\zeta_k^i$ and terminate. Else: $i \gets i+1$ and go to Step 2.
\end{algorithmic}
\end{algorithm}

As guaranteed by Theorem~\ref{Theorem2}, Algorithm~\ref{algorithm1} generates a sequence of estimates $\{\zeta_k^i\}_{i\geq0}$ which is non-decreasing in its likelihoods.

Although the densities $V_k$ and $W_k$ are assumed to be known, the integral (\ref{Q_int_Alg}) in Algorithm~\ref{algorithm1} requires the knowledge of $p(\zeta_{k-1}|\mathcal{Y}_{0:k-1})$ as well. In general, outside the linear Gaussian case, such an integral cannot be analytically evaluated. Hence, a numerical approximation has to be considered. In this paper, we turn to using a particle filter.

\subsection{EM with a particle filter}
The EM algorithm, so far, relies on the provision of the filtered density $p(\zeta_{k-1}|\mathcal{Y}_{0:k-1})$ for the calculation at the E-step. A particle filter \cite{doucet2000sequential} provides a version of this information through a set of particles, i.e. a collection of
values of $\{\tilde \zeta^j_{k-1}\}_{j=1}^N$, and possibly their corresponding importance weights $\{\omega_{k-1|k-1}^j\}_{j=1}^N$. The density is approximated by
\begin{equation}
    p(\zeta_{k-1}|\mathcal{Y}_{0:k-1}) \approx  \sum_{j=1}^N \omega_{k-1|k-1}^j\delta(\zeta_{k-1}-\tilde \zeta^j_{k-1}),
\label{part_approx}
\end{equation}
where $\delta(\cdot)$ is the delta function.
If the particles are produced by resampling then each of the importance weights $\omega_{k-1|k-1}^j=1/N$. Using these particles in (\ref{part_approx}), the expectation $Q_\zeta(\zeta_k,\zeta_k^i)$, in (\ref{Q_int_Alg}), can be replaced by its particle approximation $\hat Q^N_\zeta(\zeta_k,\zeta_k^i)$. Algorithm \ref{algorithm1} can then be recast using the particles for the E-step as follows.

\begin{algorithm}[H]
\caption{EM State Filter (EMSF) Algorithm}\label{algorithm2}
\begin{algorithmic}[1]
        \State Given $y_k$, the particles $\{\tilde \zeta^j_{k-1}\}_{j=1}^N$, their normalized importance weights $\{\omega_{k-1|k-1}^j\}_{j=1}^N$ and $\zeta_{k-1}^*$, set $i \gets 0$ and let $\zeta_k^i=f(\zeta_{k-1}^*) + \mathbb{E}w_k$;
        \State Evaluate the summation:
        \begin{align}
             \hat Q^N_\zeta(\zeta_k,\zeta_k^i)&=\sum_{j=1}^N \Big [\log V_k(y_k-g(\zeta_k))\nonumber\\&+ \log W_{k-1}(\zeta_k-f(\tilde \zeta_{k-1}^j)) \Big ]\times \nonumber \\& W_{k-1}(\zeta_k^i-f(\tilde \zeta_{k-1}^j))\omega_{k-1|k-1}^j 
        \end{align}
        \State Evaluate $\zeta_k^{i+1}$:
        \begin{equation}
            \zeta_k^{i+1} =\text{arg} \max_{\zeta_k \in \R^{r_\zeta}} \hat Q^N_\zeta(\zeta_k,\zeta_k^{i})
        \end{equation}
        \State If $\{\zeta_k^i\}_{i\geq0}$ satisfies a convergence criterion: set $\zeta_k^*=\zeta_k^i$ and terminate. Else: $i \gets i+1$ and go to Step 2.
\end{algorithmic}
\end{algorithm}

We apply EMSF in the following computed examples.
\subsection{Example 1: Linear Gaussian state-space model}
\label{LGSSM}
In the linear Gaussian state-space model case, all random variables under consideration are Gaussian. Hence, the conditional mean is equivalent to the conditional mode \cite{rauch1965maximum}. In the following example, the Kalman filter, which is designed to track the propagation of the conditional mean, is used in comparison with the EMSF. It will be shown that the EMSF, starting from a random initialization $\zeta_k^0$, converges to the Kalman Filter solution, which is guaranteed by Corollary~\ref{corollary1}.

Consider the following model,
\begin{equation}
\begin{aligned}
\zeta_{k+1} & = F 
\zeta_{k} + 
    w_k,\\
    y_{k}&=H
\zeta_{k}
+v_k,\\
v_k &\sim \mathcal{N}(0,R), \,
w_k \sim \mathcal{N}(0,S)\\
\zeta_0&\sim \mathcal N(\zeta_{0|-1},\Sigma_{0|-1})\\
    \label{eqn466}
\end{aligned}
\end{equation}
where the sequences $\{w_k\}$ and $\{v_k\}$ and the random variable $\zeta_0$ follow the independence and whiteness assumptions as in Section~\ref{section2}. The term $\mathcal N(\mu,\Sigma)$ denotes the multivariate normal density of mean $\mu$ and covariance $\Sigma$, and $\mathcal N(x|\mu,\Sigma)$ denotes the value of this density when evaluated at $x$.

We use the following particle approximation,
\begin{equation}
    p(\zeta_{k-1}|\mathcal{Y}_{0:k-1}) \approx  \sum_{j=1}^N \omega_{k-1|k-1}^j\delta(\zeta_{k-1}-\tilde \zeta^j_{k-1}).
\end{equation}
The summation in Algorithm \ref{algorithm2}, ignoring constant terms superfluous to the maximization, can be simplified to,
 \begin{align}
             \hat Q^N_\zeta(\zeta_k,\zeta_k^i)&=\sum_{j=1}^N \Big [-(y_k-H\zeta_k)^TR^{-1}(y_k-H\zeta_k) \nonumber\\&\hskip 5mm -(\zeta_k-F\tilde \zeta_{k-1}^j)^TS^{-1}(\zeta_k-F\tilde \zeta_{k-1}^j) \Big] \nonumber\\&\hskip 5mm \times \lambda^j\omega_{k-1|k-1}^j \label{LGSS_Ex}
\end{align}
where $\lambda^j=W_{k-1}(\zeta_k^i-F\tilde \zeta_{k-1}^j)=\mathcal{N}(\zeta_k^i-F\tilde \zeta_{k-1}^j|0,S)$. Since (\ref{LGSS_Ex}) is quadratic in $\zeta_k$, the maximizer $\zeta_k^{i+1}$ can be found analytically,
\begin{align*}
    \zeta_{k}^{i+1} \text{ is such that } \nabla_{\zeta_k}Q^N_\zeta(\zeta_k^{i+1},\zeta_k^i)=0.
\end{align*}
Hence,
\begin{align}
    \zeta_k^{i+1}&=\Big [ H^TR^{-1}H+S^{-1}\Big]^{-1} \frac{1}{\sum_{l=1}^N\lambda^{l}\omega_{k-1|k-1}^{l}} \times\nonumber \\&\Big \{ H^TR^{-1}y_k+S^{-1}F\sum_{j=1}^N\tilde \zeta_{k-1}^j\lambda^j\omega_{k-1|k-1}^j\Big \},\label{eq:infofilt}
\end{align}
or, using the matrix inversion lemma,
\begin{align}
    \zeta_k^{i+1}&=\Big [ F\sum_{j=1}^N\tilde \zeta_{k-1}^j\lambda^j\omega_{k-1|k-1}^j + \nonumber \\& \hskip -10mm\mathcal{B}(y_k-HF\sum_{j=1}^N\tilde \zeta_{k-1}^j\lambda^j\omega_{k-1|k-1}^j) \Big] \frac{1}{\sum_{l=1}^N\lambda^{l}\omega_{k-1|k-1}^{l}} \label{EMSFLG}
\end{align}
with
\begin{align*}
    \mathcal{B}=SH^T(HSH^T+R)^{-1}.
\end{align*}
Note that, while \eqref{EMSFLG} bears a passing resemblance to the Kalman filter recursion as \eqref{eq:infofilt} does to the information filter, the $\lambda^j$ terms and $\hat Q^N_\zeta(\zeta_k,\zeta^i_k)$ in \eqref{LGSS_Ex} are dependent on the previous iterate $\zeta^i_k$ and the update follows the maximizer.

Now take 
\begin{align*}
    &\zeta_k=\begin{bmatrix}\zeta_{k,1}\\\zeta_{k,2}\\\zeta_{k,3}\end{bmatrix},\;
    F=
    \begin{bmatrix}
    0.66&-1.31&-1.11\\
    0.07& 0.73&-0.06\\
    0.00&0.08&0.80
    \end{bmatrix},\\ &H=
    \begin{bmatrix}
    0&1&1
    \end{bmatrix},
    \,
    S=
    \begin{bmatrix}
    0.2&0&0\\
    0&0.3&0\\
    0&0&0.5
    \end{bmatrix},\,
    R=0.1,\\
    &\zeta_{0|-1}=\begin{bmatrix}0\\0\\\end{bmatrix},\;\;
    \Sigma_{0|-1}=
    \begin{bmatrix}
     0.3&0&0\\
     0&0.3&0\\
     0&0&0.3
    \end{bmatrix}.
\end{align*}

\begin{figure}[h]
\centering 
\includegraphics[width=3in,height=4.0in]{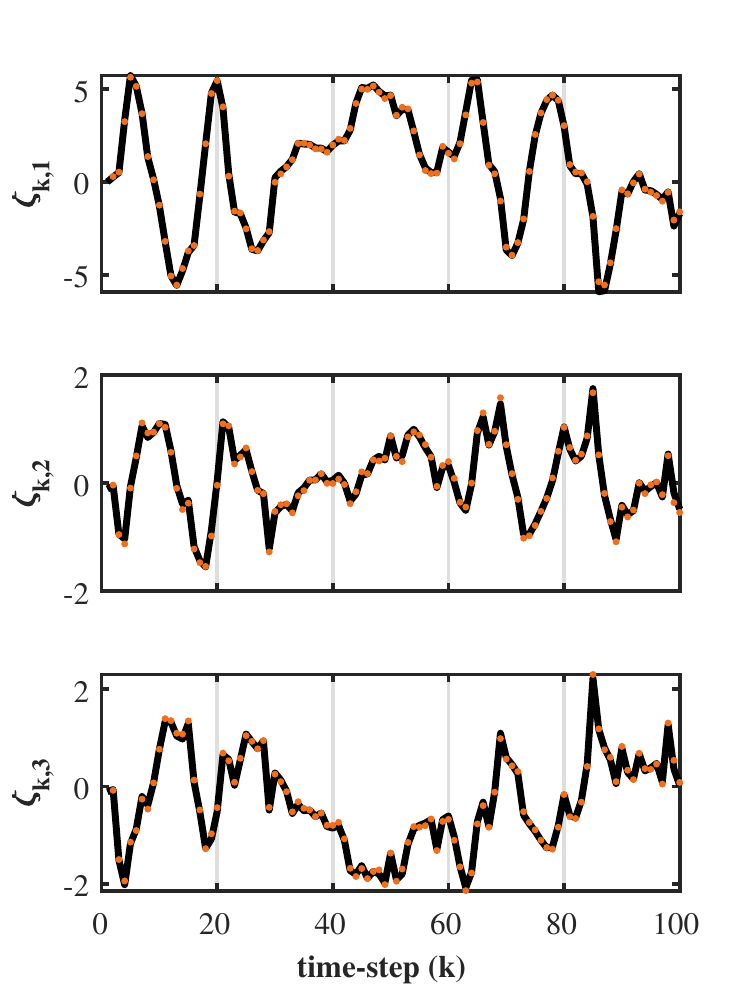} \caption{The Kalman Filter solution in black, and the EMSF solution in orange dots.\label{fig:eg1}}
\end{figure}

\begin{figure}[h]
\centering 
\includegraphics[width=3in,height=1.5in]{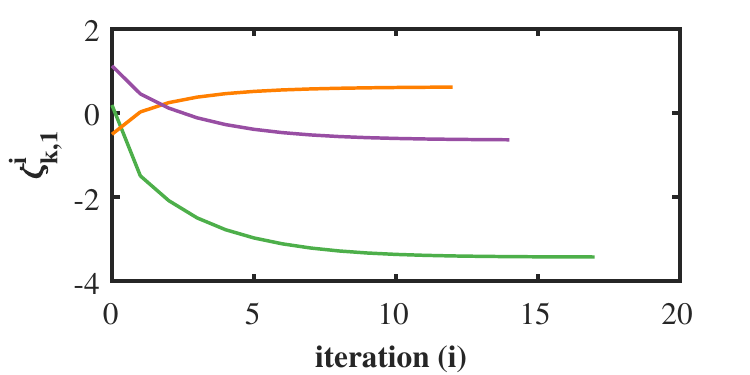} \caption{The convergence of the sequences $\{\zeta_{k,1}^i:i\geq 0\}$ for $k=15,55,90$ in green, orange and purple, respectively.\label{fig:eg1b}}
\end{figure}

A simulation over $100$ time-steps, with $N=2000$ particles, is conducted with the results plotted in Figure~\ref{fig:eg1}. The EMSF recursion at each time-step $k$, as described by (\ref{EMSFLG}), is initialized by $\zeta_k^0\sim \mathcal{N}(0,I_{3\times3})$. The results show that the EMSF converges to the Kalman Filter solution, which property follows from Corollary~\ref{corollary1}.
Figure~\ref{fig:eg1b}, depicts the convergence of the iterates at three individual times of $\zeta_{k,1}$, where the algorithm stops whenever the maximum absolute relative error is below $0.005$, i.e. $\max\,_{q\in\{1,2,3\}} | (\zeta_{k,q}^{i+1}-\zeta_{k,q}^i)/\zeta_{k,q}^{i+1}| \leq 0.5 \%$. 

The EMSF algorithm is of $\mathcal{O}(N)$ computational complexity, per iteration, with $N$ the number of particles. Ten simulations were conducted using \matlab, an uncompiled interpretive program, on a computer with a 2.5 GHz Intel Core i5-7200U processor and 8.00 GB of RAM. The running times for the one hundred time points averaged over the ten runs were: 3 milliseconds for the Kalman filter, 3.4 seconds for the bootstrap particle filter, and 26.4 seconds for EMSF including the bootstrap particle filter.
  
\subsection{Example 2: Nonlinear state-space model with
multi-modal/skewed state distribution}\label{section2E}
In this example, the EMSF is tested with a nonlinear state-space model with highly skewed and/or multimodal state densities. The EMSF solution is compared to the conditional mean of the bootstrap particle filter. This example also serves as an introduction to the next section, in which the EMSF is shown to reduce to a convergent fixed-point iteration for an important class of nonlinear systems.

\begin{figure}[h]
\centering 
\includegraphics{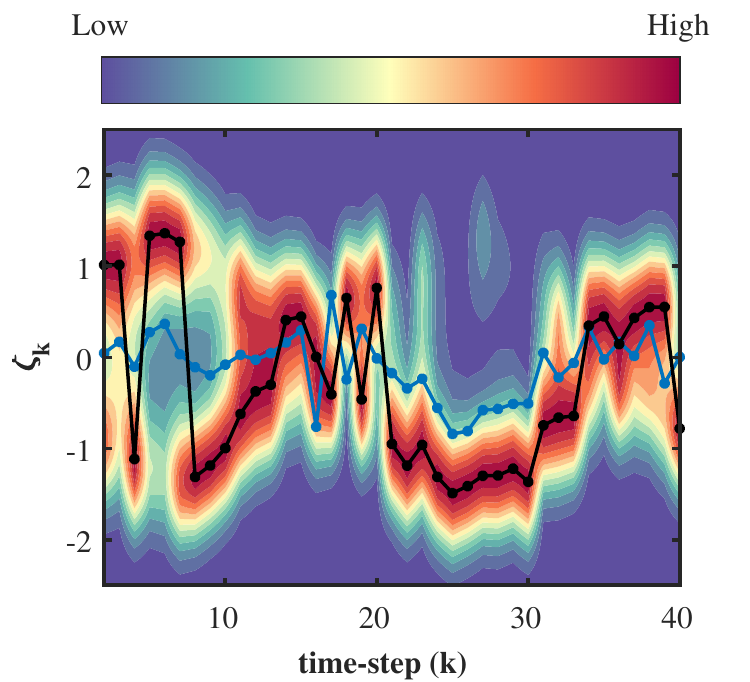} \caption{The shading represents the empirical density function (up to a positive multiplicative constant). In black is the output of the EMSF algorithm, and in blue is the conditional mean of the particle filter.\label{fig:eg2}}
\end{figure}

\begin{figure}[h]
\centering 
\includegraphics[width=3.0in,height=2.0in]{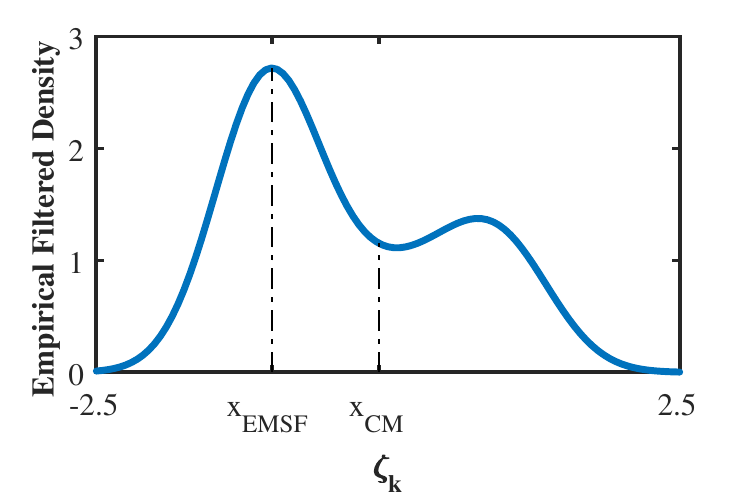} \caption{The bi-modal empirical density function of $\zeta_k$ at $k=10$ (apart from a multiplicative constant). x$_{EMSF}$ is the solution of the EMSF algorithm and x$_{CM}$ is the conditional mean of the bootstrap particle filter.\label{fig:eg2b}}
\end{figure}

\begin{figure}[h]
\centering 
\includegraphics[width=3.0in,height=2.0in]{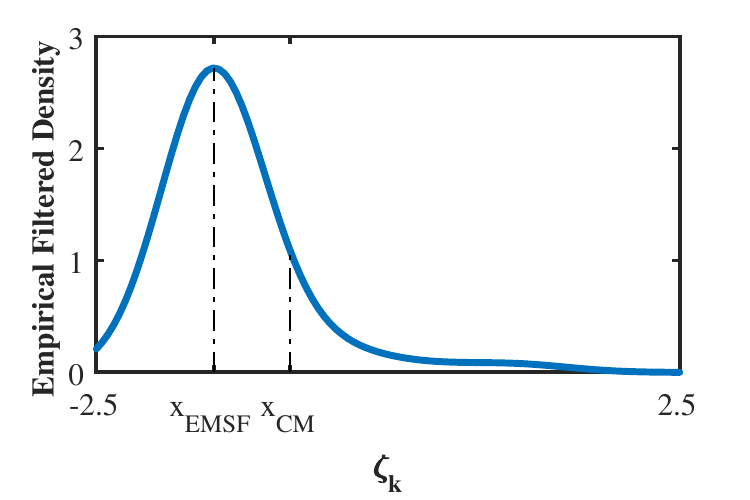} \caption{The highly skewed empirical density function of $\zeta_k$ at $k=25$. x$_{EMSF}$ is the solution of the EMSF algorithm and x$_{CM}$ is the conditional mean of the bootstrap particle filter.\label{fig:eg2c}}
\end{figure}

Consider the following nonlinear state-space model,
\begin{equation}
    \begin{aligned}
    \zeta_{k+1}&=\alpha_k\tanh{(\pi \zeta_k)}+w_k\\
    y_k&=\frac{1}{2}\zeta_k+v_k\\
    w_k &\sim \mathcal{N}(0,\frac{1}{5}),\, v_k \sim \mathcal{N}(0,1)\\
    \zeta_0 &\sim \mathcal{N}(0,1)
    \end{aligned}
\end{equation}
 where $\alpha_k=(1+0.5\sin(2\pi k/20))$. The  summation (\ref{Q_int_Alg}) in  Algorithm~\ref{algorithm2},  after  ignoring constant terms, can be simplified to,
  \begin{align}
             \hat Q^N_\zeta(\zeta_k,\zeta_k^i)&=\sum_{j=1}^N \Big [-(y_k-\frac{1}{2}\zeta_k)^2-\nonumber\\&\hskip -10mm 5(\zeta_k-\alpha_{k-1}\tanh(\pi \tilde\zeta_{k-1}^j))^2 \Big]
             \times \lambda^j\omega_{k-1|k-1}^j. \label{LGSS_Ex2}
\end{align}
where $\lambda^j=\mathcal{N}(\zeta_k^i-\alpha_{k-1}\tanh(\pi \tilde\zeta_{k-1}^j) |0,1/5)$. Since (\ref{LGSS_Ex2}) is quadratic in $\zeta_k$, the unique maximizer is given by
\begin{align}
    \zeta_{k}^{i+1}&=\frac{1}{10.5} \frac{1}{\sum_{l=1}^N \lambda^l\omega_{k-1|k-1}^l} \times \nonumber \\& \sum_{l=1}^N \big [ y_k + 10\alpha_k \tanh{(\pi \tilde\zeta_{k-1}^j)} \lambda^j\omega_{k-1|k-1}^j\big ]. \label{recursionNon}
\end{align}
The simulation is carried out over $40$ time-steps, and $N=500$ particles. The EMSF, at each time $k$, is defined by the recursion (\ref{recursionNon}) and is initialized by $5$ independent initial guesses $\zeta_k^0\sim \mathcal{U}[-2,2]$. The number of iterations is set to be $\leq 10$, and the point of highest final likelihood is taken over the five initial conditions. Notice that
\begin{align*}
     p(\zeta_k| \mathcal{Y}_{0:k})&=p(\zeta_k| y_k,\mathcal{Y}_{0:k-1})\\
    &\propto  p(y_k|\zeta_k).p(\zeta_k|\mathcal{Y}_{0:k-1})\\
    &=p(y_k|\zeta_k) \int p(\zeta_k|\zeta_{k-1}) p(\zeta_{k-1}|\mathcal{Y}_{0:k-1})\,d\zeta_{k-1}.
\end{align*}
By using the particle approximation of the filtered density $p(\zeta_{k-1}|\mathcal{Y}_{0:k-1})$, we have
\begin{align*}
    p(\zeta_k| \mathcal{Y}_{0:k})&\approx  p(y_k|\zeta_k) \sum_{j=1}^N p(\zeta_k|\tilde \zeta_{k-1}^j)\omega_{k-1|k-1}^j
\end{align*}
which is the empirical filtered density, and can be used to compare the likelihoods of the convergence points resulting from multiple initial guesses. For this example,
\begin{align*}
    p(\zeta_k| \mathcal{Y}_{0:k})&\approx\mathcal{N}(y_k-\frac{1}{2}\zeta_k|0,1)\times\\& \sum_{j=1}^N \mathcal{N}(\zeta_k-\alpha_{k-1}\tanh(\pi\tilde \zeta_{k-1}^j)|0,\frac{1}{5})\omega_{k-1|k-1}^j
\end{align*}
where the particles $\{\tilde \zeta_{k-1}^j\}$ and their weights $\{\omega_{k-1|k-1}^j\}$ follow the definition as in the EMSF algorithm. The results are shown in Figures~\ref{fig:eg2} to \ref{fig:eg2c}.


In this example, the EMSF reduces to a fixed-point iteration. This leads to the following result for state-space models with Gaussian noises and linear measurement equations, in which this simplification of EMSF is possible.

\subsection{Gaussian state-space models with linear measurement equations}\label{section2F}
In the case of a nonlinear state-space model with linear measurement equation and affine Gaussian noises, Algorithm~\ref{algorithm2} reduces to a fixed-point iteration method which is convergent to a stationary point of the empirical filtered density.

\vskip3mm
\begin{theorem} \label{theorem3}
For the nonlinear state-space system (\ref{NonSys}) defined in Section~\ref{section2} with $g(\zeta_k)=H\zeta_k$ for $H\in\mathbb R^{r_y\times r_\zeta}$ and $w_k\sim\mathcal{N}(0,S_k)$, $v_k\sim\mathcal{N}(0,R_k)$, Steps~2 and 3 of Algorithm~\ref{algorithm2} reduce to the following fixed-point iteration.
\begin{align*}
    \zeta_k^{i+1}&= \frac{1}{\sum_{l=1}^N\lambda^{l}\omega_{k-1|k-1}^{l}} \times \Big [\sum_{j=1}^Nf(\tilde \zeta_{k-1}^j)\lambda^j\omega_{k-1|k-1}^j+\\
    &\hskip 10mm\mathcal{B}_k(y_k-H\sum_{j=1}^Nf(\tilde \zeta_{k-1}^j)\lambda^j\omega_{k-1|k-1}^j)\Big ],
\end{align*}
where
\begin{align*}
    \mathcal{B}_k=S_{k-1}H^T(HS_{k-1}H^T+R_k)^{-1},
\end{align*}
and
\begin{align*}
\lambda^j=\mathcal{N}(\zeta_k^i-f(\tilde \zeta_{k-1}^j)|0,S_{k-1}), \quad j=1,2,\hdots N.
\end{align*}
\end{theorem}
\textit{Proof.}~
The summation (\ref{Q_int_Alg}) in Algorithm~\ref{algorithm2} reduces to
\begin{align*}
             \hat Q^N_\zeta(\zeta_k,\zeta_k^i)&=\sum_{j=1}^N \Big [-(y_k-g(\zeta_k))^TR_k^{-1}(y_k-g(\zeta_k))\nonumber\\&\hskip -15mm -(\zeta_k-f(\tilde \zeta_{k-1}^j))^TS_{k-1}^{-1}(\zeta_k-f(\tilde \zeta_{k-1}^j)) \Big ]\times \nonumber \lambda^j\omega_{k-1|k-1}^j 
\end{align*}
which is quadratic in $\zeta_k$. Hence, the unique maximizer of $Q^N_\zeta(\zeta_k,\zeta_k^i)$ can be found by setting $\nabla_{\zeta_k}Q^N_\zeta(\zeta_k,\zeta_k^i)$ to zero. Thus,
\begin{align*}
              \zeta_k^{i+1}&=\Big [ H^TR_k^{-1}H+S_{k-1}^{-1}\Big]^{-1} \frac{1}{\sum_{l=1}^N\lambda^{l}\omega_{k-1|k-1}^{l}} \times\nonumber \\&\Big \{ H^TR_k^{-1}y_k+S_{k-1}^{-1}\sum_{j=1}^Nf(\tilde \zeta_{k-1}^j)\lambda^j\omega_{k-1|k-1}^j\Big \}.
        \end{align*}
Applying the matrix inversion lemma, as in Section~\ref{LGSSM}, completes the proof.
\hfill $\square$

We have the following ML state filtering algorithm.
\begin{algorithm}[H]
\caption{Fixed-Point State Filter (FPSF) Algorithm for linear measurements and affine Gaussian noises}\label{algorithmFP}
\begin{algorithmic}[1]
\State Subject to the assumptions of Theorem~\ref{theorem3} and 
 given: $y_k$, the particles $\{\tilde \zeta^j_{k-1}\}_{j=1}^N$, their normalized importance weights $\{\omega_{k-1|k-1}^j\}_{j=1}^N$ and the initial guess $\zeta_k^0$, set $i~\gets~0$;
        \State Do:
        \begin{align*}
              \zeta_k^{i+1}&= \frac{1}{\sum_{l=1}^N\lambda^{l}\omega_{k-1|k-1}^{l}} \times \Big [\sum_{j=1}^Nf(\tilde \zeta_{k-1}^j)\lambda^j\omega_{k-1|k-1}^j+\\
    &\hskip 12mm\mathcal{B}_k(y_k-H\sum_{j=1}^Nf(\tilde \zeta_{k-1}^j)\lambda^j\omega_{k-1|k-1}^j)\Big ]
        \end{align*}
        where
\begin{align*}
    \mathcal{B}_k=S_{k-1}H^T(HS_{k-1}H^T+R_k)^{-1}
\end{align*}
and
\begin{align*}
\lambda^j=\mathcal{N}(\zeta_k^i-f(\tilde \zeta_{k-1}^j)|0,S_{k-1}), \quad j=1,2,\hdots N
\end{align*}
        \State If $\{\zeta_k^i\}_{i\geq0}$ satisfies a convergence criterion, terminate. Else: $i~\gets~i+1$ and go to step 2.
\end{algorithmic}
\end{algorithm}
\section{ML state prediction by the expectation maximization algorithm}\label{section4}

Given,
\begin{itemize}
    \item \vskip -3mm the problem statement and assumptions as in Section~\ref{section2},
    \item the particle approximation of the time-$m$ filtered density
    $p(\zeta_{m}|\mathcal{Y}_{0:m})$, for $m\in\{1,2,\hdots,k-1\}$, i.e.
    \begin{equation*}
    p(\zeta_{m}|\mathcal{Y}_{0:m}) \approx 
    \sum_{j=1}^N \omega_{m|m}^j\delta(\zeta_{m}-\tilde \zeta^j_{m|m})
\label{part_approx_p}    \end{equation*}
    \item initial state estimate $\zeta^0_k$,
\end{itemize}
the time-update of the bootstrap particle filter can be used to propagate the particles and achieve the particle approximation
\begin{align}
     p(\zeta_{k-1}|\mathcal{Y}_{0:m}) &\approx 
    \sum_{j=1}^N \omega_{m|m}^j\delta(\zeta_{k-1}-\tilde \zeta^j_{k-1|m}). \label{part_approx_pred}
\end{align}
The importance weights are unchanged because no extra measurements are available after time-$m$. The EM algorithm can then be used to evaluate the MLE of $\zeta_k$ of the predicted density $p(\zeta_k|\mathcal{Y}_{0:m})$, using the particle approximation (\ref{part_approx_pred}). The E-step is performed by evaluating the following expectation
\begin{align*}
     &Q^p_\zeta(\zeta_k,\zeta_k^i)\nonumber\\
     &=\E_{\zeta_{k-1}|\zeta_k^{i},\mathcal{Y}_{0:m}} \{\log p(\zeta_k|\zeta_{k-1},\mathcal{Y}_{0:m})|\zeta_k^{i},\mathcal{Y}_{0:m}\},\nonumber\\
     &=\int \Big \{\log p(\zeta_k|\zeta_{k-1},\mathcal{Y}_{0:m})\nonumber\\&\hskip 10mm p(\zeta_{k-1}|\zeta_k^{i},\mathcal{Y}_{0:m
     })\Big \}\,d\zeta_{k-1},
\end{align*}
which, analogous to the proof of Theorem~\ref{theorem1}, is equivalent to
\begin{align}
    Q^p_\zeta(\zeta_k,\zeta_k^i)
    &=\int \Big \{\big [\log W_{k-1}(\zeta_k-f(\zeta_{k-1})) \big ]\times \nonumber \\&\hskip -10mm W_{k-1}(\zeta_k^i-f(\zeta_{k-1}))p(\zeta_{k-1}|\mathcal{Y}_{0:m})\Big \}\,d\zeta_{k-1}.\label{Q_int_p}
\end{align}

The particle approximation (\ref{part_approx_pred}) can be used to construct $Q^{p,N}_\zeta$, an approximate of (\ref{Q_int_p}) up to additive and multiplicative constants which are independent from $\zeta_k$,
\begin{align*}
             \hat Q^{p.N}_\zeta(\zeta_k,\zeta_k^i)&=\sum_{j=1}^N \Big [\log W_{k-1}(\zeta_k-f(\tilde \zeta_{k-1|m}^j)) \Big ]\nonumber \\&\times W_{k-1|m}(\zeta_k^i-f(\tilde \zeta_{k-1|m}^j))\omega_{m|m}^j.
\end{align*}
And the M-step is performed by maximizing this approximate over $\zeta_k$. Hence, the EM algorithm for ML state prediction,
\begin{algorithm}[H]
\caption{EM State Predictor (EMSP) Algorithm }\label{algorithm4}
\begin{algorithmic}[1]
        \State Given the particles $\{\tilde \zeta^j_{k-1|m}\}_{j=1}^N$, their normalized importance weights $\{\omega_{m|m}^j\}_{j=1}^N$ and $\zeta_{k-1}^*$, set $i \gets 0$ and let $\zeta_k^i=f(\zeta_{k-1}^*) + \mathbb{E}w_k$;
        \State Evaluate the summation:
        \begin{align}
             \hat Q^{p.N}_\zeta(\zeta_k,\zeta_k^i)&=\sum_{j=1}^N \Big [\log W_{k-1}(\zeta_k-f(\tilde \zeta_{k-1|m}^j)) \Big ]\nonumber \\&\times W_{k-1}(\zeta_k^i-f(\tilde \zeta_{k-1|m}^j))\omega_{m|m}^j \label{Q_sum_Alg_p}
        \end{align}
        \State Evaluate $\zeta_k^{i+1}$:
        \begin{equation}
            \zeta_k^{i+1} =\text{arg} \max_{\zeta_k \in \R^{r_\zeta}} \hat Q^{p,N}_\zeta(\zeta_k,\zeta_k^{i})
        \end{equation}
        \State If $\{\zeta_k^i\}_{i\geq0}$ satisfies a convergence criterion: set $\zeta_k^*=\zeta_k^i$ and terminate. Else: $i \gets i+1$ and go to step 2.
\end{algorithmic}
\end{algorithm}

Similar versions of Theorem~\ref{Theorem2} and Corollary~\ref{corollary1} hold for the sequence $\{\zeta_k^i\}_{i\geq0}$ generated by Algorithm~\ref{algorithm4}, i.e. the sequence $\{\log p(\zeta_k^i|\mathcal{Y}_{m})\}_{i\geq0}$ is non-decreasing, and if $\log p(\zeta_k|\mathcal{Y}_{m})$ is unimodal and its MLE is the only stationary point, then $\{\zeta_k^i\}_{i\geq0}$ of Algorithm~\ref{algorithm4} converges to the MLE.

\section{ML state smoothing by the expectation maximization algorithm}\label{section5}
In contrast to prediction and filtering, smoothed state estimation involves considerably greater complexity and, in the MLE environment, greater demands of the particle filter. This leads to an EM algorithm which has complexity $\mathcal{O}(N^2)$. The algorithm EMSS, however, follows directly, as we now show.

Given,
\begin{itemize}
    \item \vskip -3mm the problem statement and assumptions as in Section~\ref{section2},
    \item the measurements sequence $\mathcal{Y}_{0:n}$, for $n > k+1$,
    \item The particle approximations of the following densities:
\begin{align}
    \hskip -5mm p(\zeta_{k-1}|\mathcal{Y}_{0:k-1}) &\approx  \sum_{j=1}^N \omega_{k-1|k-1}^j\delta(\zeta_{k-1}-\tilde \zeta^j_{k-1|k-1}), \label{part_approx_S1}
\\
p(\zeta_{k}|\mathcal{Y}_{0:k}) &\approx  \sum_{j=1}^N \omega_{k|k}^j\delta(\zeta_{k}-\tilde \zeta^j_{k|k}),\label{part_approx_S2}\\
p(\zeta_{k+1}|\mathcal{Y}_{0:n}) &\approx  \sum_{j=1}^N \omega_{k+1|n}^j\delta(\zeta_{k+1}-\tilde \zeta^j_{k+1|n}),\label{part_approx_S3}
\end{align}
which can be supplied by a forward–backward particle smoother \cite{doucet2000sequential},
    \item initial state estimate $\zeta^0_k$,
\end{itemize}
The EM algorithm can be used to evaluate the MLE of $\zeta_k$ of the smoothed density $p(\zeta_k|\mathcal{Y}_{0:n})$. The E-step is performed by evaluating the expectation
\begin{align}
     &Q^S_\zeta(\zeta_k,\zeta_k^i)\nonumber\\
     &=\E_{\zeta_{k+1}|\zeta_k^{i},\mathcal{Y}_{0:n}} \{ \log p(\zeta_k|\zeta_{k+1},\mathcal{Y}_{0:n})|\zeta_k^{i},\mathcal{Y}_{0:n}\},\nonumber\\
     &=\int \log p(\zeta_k|\zeta_{k+1},\mathcal{Y}_{0:n})p(\zeta_{k+1}|\zeta_k^{i},\mathcal{Y}_{0:n
     }) \,d\zeta_{k+1}. \label{Q_int_S}
\end{align}
Where in (\ref{Q_int_S}), using Bayes rule 3 times and Lemma~\ref{lemma1}, the density
\begin{align*}
   p(\zeta_k|\zeta_{k+1},\mathcal{Y}_{0:n})&= \frac{p(\mathcal{Y}_{k+1:n}|\zeta_{k+1},\zeta_{k},\mathcal{Y}_{0:k})p(\zeta_k|\zeta_{k+1},\mathcal{Y}_{0:k})}{p(\mathcal{Y}_{k+1:n}|\zeta_{k+1},\mathcal{Y}_{0:k})},\\
   &= \frac{p(\mathcal{Y}_{k+1:n}|\zeta_{k+1})p(\zeta_k|\zeta_{k+1},\mathcal{Y}_{0:k})}{p(\mathcal{Y}_{k+1:n}|\zeta_{k+1})},\\
   &=p(\zeta_k|\zeta_{k+1},\mathcal{Y}_{0:k}), \\
   &=\frac{p(\zeta_{k+1}|\zeta_{k})p(\zeta_k|\mathcal{Y}_{0:k})}{p(\zeta_{k+1}|\mathcal{Y}_{0:k})},\\
   &=\frac{p(\zeta_{k+1}|\zeta_{k})p(y_k|\zeta_k)p(\zeta_k|\mathcal{Y}_{0:k-1})}{p(\zeta_{k+1}|\mathcal{Y}_{0:k})p(y_k|\mathcal{Y}_{0:k-1})}.
\end{align*}
The term $p(y_k|\mathcal{Y}_{0:k-1})$ is independent from $\zeta_k$ and leads to an additive constant in the expectation (\ref{Q_int_S}). Thus,
\begin{align}
    \log p(\zeta_k|\zeta_{k+1},\mathcal{Y}_{0:n})&\propto \log p(\zeta_{k+1}|\zeta_{k})+\log p(y_k|\zeta_k)\nonumber \\&+\log p(\zeta_k|\mathcal{Y}_{0:k-1}), \nonumber \\
    &=\log W_k(\zeta_{k+1}-f(\zeta_k))\nonumber \\& \hskip -15mm+\log V_k(y_k-g(\zeta_k)) +\log p(\zeta_k|\mathcal{Y}_{0:k-1}). \label{bayes1S}
\end{align}

The other density in (\ref{Q_int_S}), using the same conditional independence result and Bayes rule,
\begin{align}
    p(\zeta_{k+1}|\zeta_k^{i},\mathcal{Y}_{0:n})
    &=\frac{p(\zeta_k^i|\zeta_{k+1},\mathcal{Y}_{0:n})p(\zeta_{k+1}|\mathcal{Y}_{0:n})}{p(\zeta_{k}^i|\mathcal{Y}_{0:n})}, \nonumber\\
    &=\frac{p(\zeta_k^i|\zeta_{k+1},\mathcal{Y}_{0:k})p(\zeta_{k+1}|\mathcal{Y}_{0:n})}{p(\zeta_{k}^i|\mathcal{Y}_{0:n})}, \nonumber\\
    &\hskip -10mm=\frac{p(\zeta_{k+1}|\zeta_k^i)p(\zeta_k^i|\mathcal{Y}_{0:k})p(\zeta_{k+1}|\mathcal{Y}_{0:n})}{p(\zeta_{k+1}|\mathcal{Y}_{0:k})p(\zeta_{k}^i|\mathcal{Y}_{0:n})},
\end{align}
where the terms $p(\zeta_k^i|\mathcal{Y}_{0:k})$ and $p(\zeta_k^i|\mathcal{Y}_{0:n})$ are independent from $\zeta_{k+1}$ and yield a positive multiplicative constant in the expectation (\ref{Q_int_S}). Hence,
\begin{align}
     p(\zeta_{k+1}|\zeta_k^{i},\mathcal{Y}_{0:n})
    &\propto
    \frac{p(\zeta_{k+1}|\zeta_k^i)p(\zeta_{k+1}|\mathcal{Y}_{0:n})}{p(\zeta_{k+1}|\mathcal{Y}_{0:k})}, \nonumber \\
    &\hskip -10mm=\frac{W_k(\zeta_{k+1}-f(\zeta_k^i))p(\zeta_{k+1}|\mathcal{Y}_{0:n})}{p(\zeta_{k+1}|\mathcal{Y}_{0:k})}.\label{bayes2S}
\end{align}
Substituting (\ref{bayes1S}) and (\ref{bayes2S}) in (\ref{Q_int_S}), after dropping the additive and the multiplicative constants, results in
\begin{align}
    Q^S_\zeta(\zeta_k,\zeta_k^i)
    &=\log V_k(y_k-g(\zeta_k))+\log p(\zeta_k|\mathcal{Y}_{0:k-1}) + \nonumber \\ 
    &\int \Big \{ \big [\log W_{k}(\zeta_{k+1}-f(\zeta_{k})) \big ]\times \nonumber \\&\frac{W_{k}(\zeta_{k+1}-f(\zeta_{k}^i))p(\zeta_{k+1}|\mathcal{Y}_{0:n})}{p(\zeta_{k+1}|\mathcal{Y}_{0:k})} \Big \}\, d\zeta_{k+1}. \label{Q_int_S2}
\end{align}

Notice that for the construction of (\ref{Q_int_S2}), the densities $p(\zeta_k|\mathcal{Y}_{0:k-1})$, $p(\zeta_{k+1}|\mathcal{Y}_{0:n})$ and $p(\zeta_{k+1}|\mathcal{Y}_{0:k})$ are required. 
The particle approximation (\ref{part_approx_S2}) can be used to approximate $p(\zeta_{k+1}|\mathcal{Y}_{0:k})$,
\begin{align}
    p(\zeta_{k+1}|\mathcal{Y}_{0:k})&= \int p(\zeta_{k+1}|\zeta_k) p(\zeta_k|\mathcal{Y}_{0:k})\, d \zeta_k \nonumber\\
    &\approx\sum_{j=1}^N \omega_{k|k}^j W_k(\zeta_{k+1}-f(\tilde \zeta_{k|k}^j)). \label{term2S}
\end{align}
Similarly, the density $p(\zeta_k|\mathcal{Y}_{0:k-1})$ can be approximated using the particle approximation (\ref{part_approx_S1}),
\begin{align}
    \log p(\zeta_k|\mathcal{Y}_{0:k-1})&\approx \log \sum_{j=1}^N \omega_{k-1|k-1}^jW_{k-1}(\zeta_k-f(\tilde \zeta_{k-1|k-1}^j)). \label{term1S}
\end{align}
Finally, by using the particle approximations (\ref{term1S}) and (\ref{term2S}) in (\ref{Q_int_S2}), and (\ref{part_approx_S3}) to approximate the integral term in (\ref{Q_int_S2}), after neglecting additive and multiplicative constants, we get $\hat Q^{S.N}_\zeta$, the particle approximation of $Q^S_\zeta$, 
\begin{align}
 \hat Q^{S.N}_\zeta(\zeta_k,\zeta_k^i)&= \log V_k(y_k-g(\zeta_k))+ \nonumber \\
 &\hskip -15mm\log \sum_{j=1}^N \omega_{k-1|k-1}^jW_{k-1}(\zeta_k-f(\tilde \zeta_{k-1|k-1}^j)) + \nonumber \\ 
&\hskip -15mm\sum_{t=1}^N \Big [\log W_{k}(\tilde \zeta_{k+1|n}^t-f(\zeta_{k})) \Big ]\times \nonumber \\&\hskip -15mm\frac{W_{k}(\tilde \zeta_{k+1|n}^t-f(\zeta_{k}^i))}{\sum_{d=1}^N \omega_{k|k}^d W_k(\tilde \zeta_{k+1|n}^t-f(\tilde \zeta_{k|k}^d))} \omega_{k+1|n}^t.
\end{align}
The EM algorithm for ML state smoothing,
\begin{algorithm}[H]
\caption{EM State Smoother (EMSS) Algorithm }\label{algorithm6}
\begin{algorithmic}[1]
\State Given the sets of particles $\{\tilde \zeta^j_{k-1|k-1}\}_{j=1}^N$, $\{\tilde \zeta^j_{k|k}\}_{j=1}^N$, $\{\tilde \zeta^j_{k+1|n}\}_{j=1}^N$, and their corresponding normalized importance weights, set $i~\gets~0$ and let the initial guess be $\zeta_k^i$;
\State Evaluate:
\begin{align}
 \hat Q^{S.N}_\zeta(\zeta_k,\zeta_k^i)&= \log V_k(y_k-g(\zeta_k))+ \nonumber \\
 &\hskip -15mm\log \sum_{j=1}^N \omega_{k-1|k-1}^jW_{k-1}(\zeta_k-f(\tilde \zeta_{k-1|k-1}^j)) + \nonumber \\ 
&\hskip -15mm\sum_{t=1}^N \Big [\log W_{k}(\tilde \zeta_{k+1|n}^t-f(\zeta_{k})) \Big ]\times \nonumber \\&\hskip -15mm\frac{W_{k}(\tilde \zeta_{k+1|n}^t-f(\zeta_{k}^i))}{\sum_{d=1}^N \omega_{k|k}^d W_k(\tilde \zeta_{k+1|n}^t-f(\tilde \zeta_{k|k}^d))} \omega_{k+1|n}^t. \label{Q_sum_Alg_S}
\end{align}
\State Evaluate $\zeta_k^{i+1}$:
\begin{equation}
\zeta_k^{i+1} =\text{arg} \max_{\zeta_k \in \R^{r_\zeta}} \hat Q^{S,N}_\zeta(\zeta_k,\zeta_k^{i})
\end{equation}
\State If $\{\zeta_k^i\}_{i\geq0}$ satisfies a convergence criterion: set $\zeta_k^*=\zeta_k^i$ and terminate. Else: $i \gets i+1$ and go to step 2.
\end{algorithmic}
\end{algorithm}
\section{Conclusion}

This paper contributes a novel approach to incorporating the EM algorithm, along with a particle filter, into the problem of maximum likelihood state estimation. Algorithms for maximum likelihood state filtering, prediction, and smoothing are presented, together with the desirable convergence properties they inherit from the EM algorithm. Additionally, for the wide range of nonlinear state-space systems possessing linear measurement equations and affine Gaussian noises, these algorithms reduce to derivative-free optimization through fixed-point iteration. The EMSF algorithms is tested on two examples: one using a linear third-order Gaussian system for comparison with the Kalman filter and to demonstrate; the concordance of the estimates, the computational simplicity of the Kalman filter and the computational feasibility of the EMSF combined with its attendant particle filter; the other to estimate the MLE of a state density which exhibits multi-modality and strong skewness. Further work is being pursued to understand the role of MLE state estimates in constrained control, such as Model Predictive Control where linearity and Gaussian assumptions can contravene the problem formulation.

\bibliographystyle{plain} 
\bibliography{MLSEautom} 

\begin{thebibliography}{10}

\bibitem{dejong2013efficient}
David~N DeJong, Roman Liesenfeld, Guilherme~V Moura, Jean-Fran{\c{c}}ois
  Richard, and Hariharan Dharmarajan.
\newblock Efficient likelihood evaluation of state-space representations.
\newblock {\em Review of Economic Studies}, 80(2):538--567, 2013.

\bibitem{dempster1977maximum}
Arthur~P Dempster, Nan~M Laird, and Donald~B Rubin.
\newblock Maximum likelihood from incomplete data via the {EM} algorithm.
\newblock {\em Journal of the Royal Statistical Society: Series B
  (Methodological)}, 39(1):1--22, 1977.

\bibitem{doucet2000sequential}
Arnaud Doucet, Simon Godsill, and Christophe Andrieu.
\newblock On sequential {M}onte {C}arlo sampling methods for {B}ayesian
  filtering.
\newblock {\em Statistics and Computing}, 10(3):197--208, 2000.

\bibitem{kantas2015particle}
Nikolas Kantas, Arnaud Doucet, Sumeetpal~S Singh, Jan Maciejowski, Nicolas
  Chopin, et~al.
\newblock On particle methods for parameter estimation in state-space models.
\newblock {\em Statistical Science}, 30(3):328--351, 2015.

\bibitem{lange1995gradient}
Kenneth Lange.
\newblock A gradient algorithm locally equivalent to the {EM} algorithm.
\newblock {\em Journal of the Royal Statistical Society: Series B
  (Methodological)}, 57(2):425--437, 1995.

\bibitem{le2013convergence}
Sylvain Le~Corff and Gersende Fort.
\newblock Convergence of a particle-based approximation of the block online
  expectation maximization algorithm.
\newblock {\em ACM Transactions on Modeling and Computer Simulation (TOMACS)},
  23(1):1--22, 2013.

\bibitem{lehmann2006theory}
Erich~L Lehmann and George Casella.
\newblock {\em Theory of Point Estimation}.
\newblock Springer Science \& Business Media, 2006.

\bibitem{li2016auxiliary}
Baibing Li, Cunjia Liu, and Wen-Hua Chen.
\newblock An auxiliary particle filtering algorithm with inequality
  constraints.
\newblock {\em IEEE Transactions on Automatic Control}, 62(9):4639--4646, 2016.

\bibitem{li2020leverage}
Mengheng Li and Marcel Scharth.
\newblock Leverage, {A}symmetry, and {H}eavy {T}ails in the
  {H}igh-{D}imensional {F}actor {S}tochastic {V}olatility {M}odel.
\newblock {\em Journal of Business \& Economic Statistics}, pages 1--17, 2020.

\bibitem{lindholm2018learning}
Andreas Lindholm and Fredrik Lindsten.
\newblock Learning dynamical systems with particle stochastic approximation
  {EM}.
\newblock {\em arXiv preprint arXiv:1806.09548}, 2018.

\bibitem{malik2011particle}
Sheheryar Malik and Michael~K Pitt.
\newblock Particle filters for continuous likelihood evaluation and
  maximisation.
\newblock {\em Journal of Econometrics}, 165(2):190--209, 2011.

\bibitem{meng1993maximum}
Xiao-Li Meng and Donald~B Rubin.
\newblock Maximum likelihood estimation via the {ECM} algorithm: A general
  framework.
\newblock {\em Biometrika}, 80(2):267--278, 1993.

\bibitem{mihaylova2014overview}
Lyudmila Mihaylova, Avishy~Y Carmi, Fran{\c{c}}ois Septier, Amadou Gning,
  Sze~Kim Pang, and Simon Godsill.
\newblock Overview of {B}ayesian sequential {M}onte {C}arlo methods for group
  and extended object tracking.
\newblock {\em Digital Signal Processing}, 25:1--16, 2014.

\bibitem{moon1996expectation}
Todd~K Moon.
\newblock The expectation-maximization algorithm.
\newblock {\em IEEE Signal Processing Magazine}, 13(6):47--60, 1996.

\bibitem{pitt1999filtering}
Michael~K Pitt and Neil Shephard.
\newblock Filtering via simulation: Auxiliary particle filters.
\newblock {\em Journal of the American Statistical Association},
  94(446):590--599, 1999.

\bibitem{poyiadjis2011particle}
George Poyiadjis, Arnaud Doucet, and Sumeetpal~S Singh.
\newblock Particle approximations of the score and observed information matrix
  in state space models with application to parameter estimation.
\newblock {\em Biometrika}, 98(1):65--80, 2011.

\bibitem{rauch1965maximum}
Herbert~E Rauch, F~Tung, and Charlotte~T Striebel.
\newblock Maximum likelihood estimates of linear dynamic systems.
\newblock {\em AIAA Journal}, 3(8):1445--1450, 1965.

\bibitem{schon2011system}
Thomas~B Sch{\"o}n, Adrian Wills, and Brett Ninness.
\newblock System identification of nonlinear state-space models.
\newblock {\em Automatica}, 47(1):39--49, 2011.

\bibitem{septier2009tracking}
Fran{\c{c}}ois Septier, Sze~Kim Pang, Simon Godsill, and Avishy Carmi.
\newblock Tracking of coordinated groups using marginalised {MCMC}-based
  particle algorithm.
\newblock In {\em 2009 IEEE Aerospace Conference}, pages 1--11. IEEE, 2009.

\bibitem{vankov2019filtering}
Emilian~R Vankov, Michele Guindani, Katherine~B Ensor, et~al.
\newblock Filtering and estimation for a class of stochastic volatility models
  with intractable likelihoods.
\newblock {\em Bayesian Analysis}, 14(1):29--52, 2019.

\bibitem{wei1990monte}
Greg~CG Wei and Martin~A Tanner.
\newblock A {M}onte {C}arlo implementation of the {EM} algorithm and the poor
  man's data augmentation algorithms.
\newblock {\em Journal of the American Statistical Association},
  85(411):699--704, 1990.

\bibitem{wu1983convergence}
CF~Jeff Wu et~al.
\newblock On the convergence properties of the {EM} algorithm.
\newblock {\em The Annals of Statistics}, 11(1):95--103, 1983.

\end{thebibliography}
\end{document}